\documentclass[sigconf]{acmart}
\usepackage{fancyhdr}
\usepackage{cleveref}
\usepackage{amsmath}
\usepackage{xspace}
\usepackage{amssymb}
\usepackage{color}
\usepackage{float}
\usepackage{stfloats}
\usepackage{booktabs}
\usepackage{siunitx}
\usepackage{booktabs}
\usepackage{siunitx}
\usepackage{algorithm}
\usepackage{algorithmic}
\usepackage{booktabs}
\usepackage{newfloat}
\usepackage{listings}
\AtBeginDocument{%
  }

\copyrightyear{2025} 
\acmYear{2025} 
\setcopyright{rightsretained} 
\acmConference[CHI '25]{The ACM CHI conference on Human Factors in Computing Systems is the premier international conference of Human-Computer Interaction}{April 26-May 1, 2025}{Yokohama, Japan}
\acmBooktitle{Proceedings of the 2025 CHI Conference on Human Factors in Computing Systems (CHI '25), April 26-May 1, 2025, Yokohama, Japan}
\acmDOI{XXXXXXX}
\acmISBN{979-8-XXXXXX}

\begin{document}

\newcommand{\TODO}[1]{\commentText{{\color{red}[\textbf{\textsc{TODO}}: \textit{#1}]}}}

\title{FIP: Endowing Robust Motion Capture on Daily Garment by Fusing Flex and Inertial Sensors}


\author{Jiawei Fang}
\email{jiaweif@stu.xmu.edu.cn}
\orcid{0000-0003-4745-9305}
\authornotemark[1]

\affiliation{%
  \institution{Xiamen University}
  \city{Xiamen}
  \country{China}
}

\author{Ruonan Zheng}
\email{30920241154558@stu.xmu.edu.cn}
\orcid{0009-0008-9363-3881}
\authornote{Both authors contributed equally to this research.}
\affiliation{%
  \institution{Xiamen University}
  \city{Xiamen}
  \country{China}
}

\author{Yuan Yao}
\email{furtheryao@stu.xmu.edu.cn}
\orcid{0000-0003-4363-1294}
\affiliation{%
  \institution{Xiamen University}
  \city{Xiamen}
  \country{China}
}

\author{Xiaoxia Gao}
\email{gaoxiaoxia@stu.xmu.edu.cn}
\orcid{0009-0000-1072-3542}
\affiliation{%
  \institution{Xiamen University}
  \city{Xiamen}
  \country{China}
}

\author{Chengxu Zuo}
\email{zuochengxu@stu.xmu.edu.cn}
\orcid{0000-0003-2054-2010}
\affiliation{%
  \institution{Xiamen University}
  \city{Xiamen}
  \country{China}
}

\author{Shihui Guo}
\email{guoshihui@xmu.edu.cn}
\orcid{0000-0002-1473-297X}
\authornote{Corresponding Author.}
\affiliation{%
  \institution{Xiamen University}
  \city{Xiamen}
  \country{China}
}

\author{Yiyue Luo}
\email{yiyueluo@mit.edu}
\orcid{0009-0008-5127-0496}
\affiliation{%
  \institution{MIT CSAIL}
  \city{Cambridge}
  \state{Massachusetts}
  \country{USA}
}

\renewcommand{\shortauthors}{Fang et al.}

\begin{teaserfigure}
\centering
\includegraphics[width=1.0\linewidth]{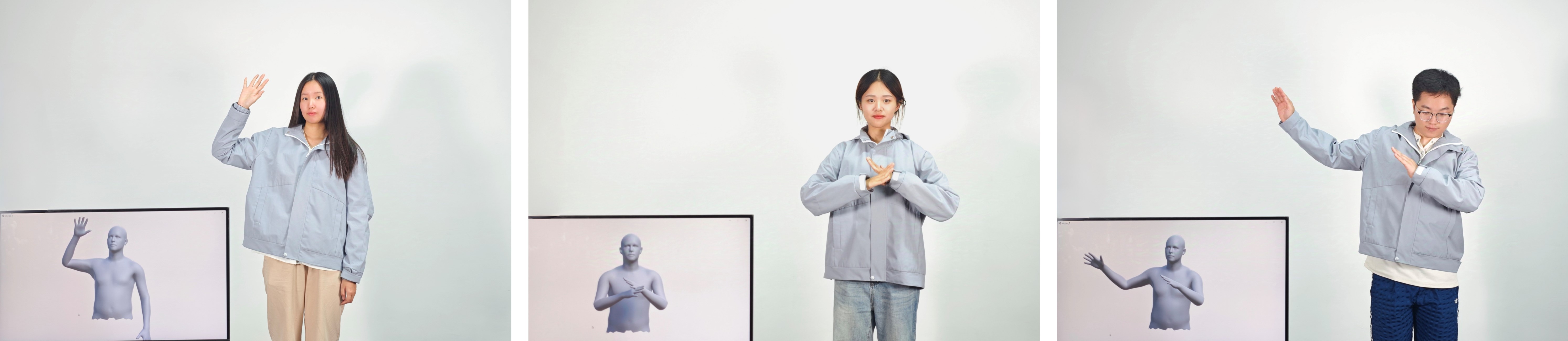}
\caption{FIP endows real-time, accurate motion capture on daily clothes by fusing flex and inertial sensors.}
\label{fig:teaser}
\end{teaserfigure}


\begin{abstract}
What if our clothes could capture our body motion accurately? This paper introduces Flexible Inertial Poser (FIP), a novel motion-capturing system using daily garments with two elbow-attached flex sensors and four Inertial Measurement Units (IMUs). To address the inevitable sensor displacements in loose wearables which degrade joint tracking accuracy significantly, we identify the distinct characteristics of the flex and inertial sensor displacements and develop a Displacement Latent Diffusion Model and a Physics-informed Calibrator to compensate for sensor displacements based on such observations, resulting in a substantial improvement in motion capture accuracy. We also introduce a Pose Fusion Predictor to enhance multimodal sensor fusion. Extensive experiments demonstrate that our method achieves robust performance across varying body shapes and motions, significantly outperforming SOTA IMU approaches with a 19.5\% improvement in angular error, a 26.4\% improvement in elbow angular error, and a 30.1\% improvement in positional error. FIP opens up opportunities for ubiquitous human-computer interactions and diverse interactive applications such as Metaverse, rehabilitation, and fitness analysis. Our project page can be seen at \href{https://fangjw-0722.github.io/FIP.github.io/}{\textcolor[rgb]{0.40, 0.21, 0.60}{\textit{\textbf{Flexible Inertial Poser}}}}.


\end{abstract}





\maketitle


\section{Introduction}

Human motion capture (MoCap) has emerged as a pivotal technology in fields such as animation, movie production, and virtual/augmented reality. Wearable motion capture offers advantages in portability, privacy friendliness, and robustness against extreme lighting and occlusion, compared to marker and vision-based approaches \cite{zhang2023textile, luo2023intelligent}. Inertial and flex sensors become the \textit{de facto} SOTA sensing mechanism that people use for wearable MoCap. Inertial sensors, also known as Inertial Measurement Units (IMUs), provide an efficient and compact means of measuring the orientation and acceleration of the human skeleton with the leading advantage of accuracy. Recent works achieve posture estimation with a sparse number (3-6) of IMUs \cite{PIPCVPR2022,yi2021transpose,jiang2022transformer,huang2018deep,yi2024physical}. However, this method requires IMUs to be tightly attached to the body for stable measurement, which inevitably produces an uncomfortable wearing experience, and results in significant measurement errors due to the underconstrained nature of joints such as elbows and knees. 

Flex sensors, also known as bending sensors, accurately measure the degree of flexion in elbow and knee joints \cite{fang2024suda, chen2023dispad, zuo2023self} by leveraging diverse sensing mechanisms, such as resistive, capacitive, and inductive sensing. They are lightweight, biocompatible, and flexible \cite{nag2017wearable, samper2020efficient}, offering solutions to the limitations posed by sparse IMU setups and enhancing measurement accuracy without compromising user comfortableness. This motivates the fusion of flex sensors and IMUs in daily garments for \textit{Clothes-based MoCap}, harnessing the strengths of each to create a more effective and comfortable MoCap solution.

However, \textit{Clothes-based MoCap} must overcome the critical sensor displacements challenges \cite{banos2012benchmark, banos2014dealing} before it can be effectively used for real-world human-computer interaction (HCI) applications \cite{zuo2024loose}. We observe two main sensor displacements arising from the loose-fitting clothing: \textit{Primary Displacement} occurs when users first put on the clothing, causing significant position shift relative to the human body due to substantial body-fabric interaction whereas \textit{Real-time Displacement} happens as the sensors move and vibrate with the loose-fitting clothing. Generally, the performance of flex sensors is significantly affected by \textit{Primary Displacement}. In contrast, \textit{Inertial-based MoCap} usually mitigates the Primary Displacement through T-pose calibration \cite{robert2017accuracy, yi2021transpose}, leaving \textit{Real-time Displacement} the dominant factor.

To this end, this paper introduces Flexible Inertial Poser (FIP), a \textit{Clothes-based MoCap} solution that achieves both robust and comfortable pose estimation with multi-modal sensor fusion (Fig. \ref{fig:teaser}). By integrating four IMUs and two auxiliary flex sensors into a loose-fitting garment, FIP leverages the unique advantages of both sensor types, offering significant advancement in accuracy, accessibility, and comfort. To address the sensor displacements based on each unique sensor type, FIP includes three components: 1) a \textbf{Displacement Latent Diffusion Model} to synthesize sensor disturbance in various conditions, addressing the \textit{Real-time Displacement} of IMUs; 2) a \textbf{Physics-informed Calibrator} to register data in different wearing condition, addressing the \textit{Primary Displacement} of flex sensors; 3) a \textbf{Pose Fusion Predictor} to fuse multi-modal sensor readings for pose estimation. 
To validate our approach, we collected a real-world testing dataset from ten users with ten different expressive motions. Extensive experimental results demonstrated the effectiveness of our method. Our work represents a beginning in exploring multi-modal sensor fusion in loose clothing, paving the way for diverse interactive applications.

In conclusion, our main contributions are as follows:
\begin{itemize}
    \item  We develop a real-time, accurate method for human motion capture using loosely worn clothes embedded with a sparse number of IMUs and auxiliary flex sensors, ensuring both user comfort and high accuracy.
    \item  We implement a novel \textbf{Displacement Latent Diffusion Model} for synthesizing inertial sensor data with \textit{Real-time Displacement}, complemented by a \textbf{Physics-Informed Calibrator} that effectively addresses the \textit{Primary Displacement} in flex sensors. In addition, a \textbf{Pose Fusion Predictor} to fuse multi-modal sensor readings for pose estimation.
    \item We evaluate our approach, which significantly reduces elbow tracking errors and achieves SOTA performance in overall joint motion capture, surpassing all competing approaches. 
    \item  We demonstrate our approach in Metaverse, rehabilitation, and fitness analysis, highlighting the benefits and potentials with \textit{Clothes-based MoCap}. 
\end{itemize}

\section{Related Work}
\subsection{Motion Capture}
Many applications such as gaming, bio-mechanical analysis, movie production, and emerging human-computer interaction paradigms such as Virtual and Augmented Reality (VR/AR) require a means to human motion capture (MoCap). Such applications impose three challenging constraints on pose reconstruction: (i) precise MoCap in real-time, (ii) work in everyday settings including in/outdoors and easy to access, (iii) minimally intrusive in terms of user experience \cite{zhou2021monocular, yi2021transpose, moeslund2006survey}.

\begin{figure*}[ht]
\centering
\includegraphics[width=1.0\linewidth]{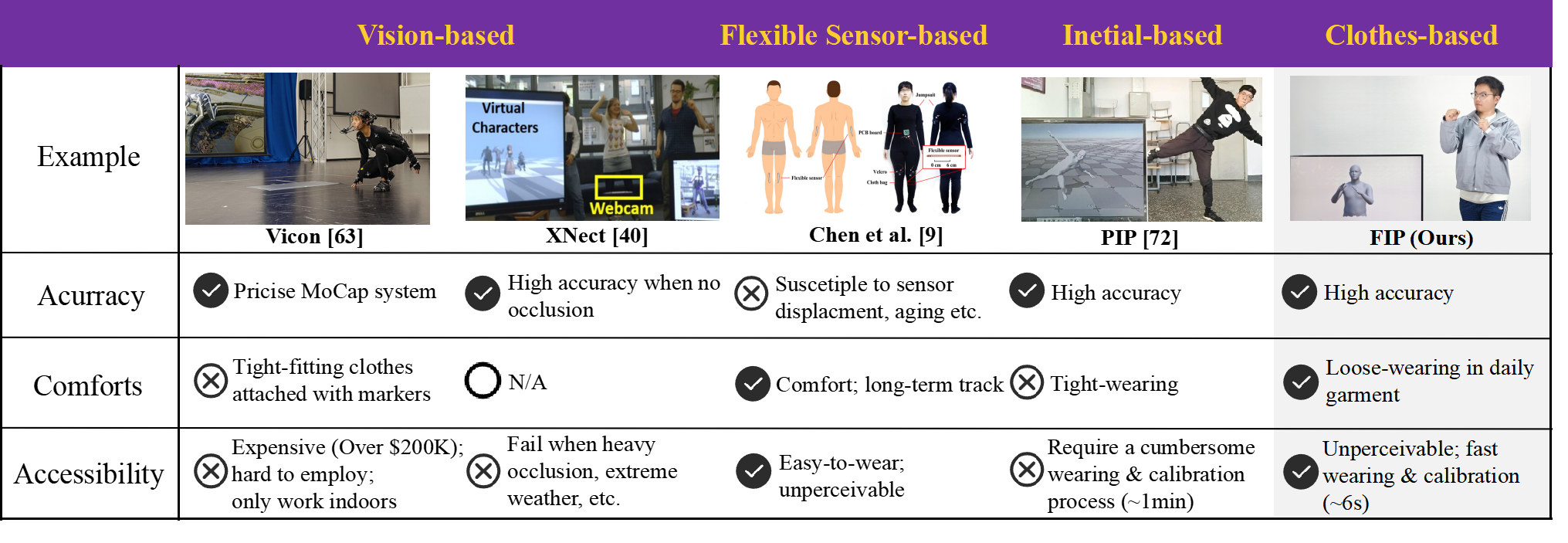}
\caption{Comparison of related MoCap methods.}
\label{fig:related}
\end{figure*}

We conducted a detailed discussion on the advantages and disadvantages of four different paradigms of MoCap methods, as outlined below, and summarized our findings in Fig. \ref{fig:related}. In summary, previous methods relied on a single type of sensor, limiting the potential for accurate motion capture. In contrast, our method, FIP, integrates both IMUs and flex sensors into loose-fitting clothing, achieving both accuracy and comfort in motion capture.

\subsubsection{Vision-based MoCap} Most commonly, the MoCap task is achieved using commercial marker-based systems like 
Vicon \cite{vicon_website}. These systems \cite{optiTrack_website, chatzitofis2021democap, ghorbani2021soma} achieve high accuracy by tracking markers on the body. However, they require costly infrastructure and physical markers placed on users, limiting accessibility in everyday contexts. Markerless single RGB or RGB-D cameras methods have gained attraction \cite{alldieck2018video, lin2023design, saini2023smartmocap, xu2022vitpose, li2022mhformer, shin2024wham, feng2023diffpose, lu2024rtmo, you2023co, dwivedi2024tokenhmr, ye2023distilpose, yoshiyasu2023deformable, xu2023vitpose++, yu2021function4d, pesavento2024anim, lu20233d, mehta2020xnect}, but similar to multi-camera systems \cite{zhang2021direct, wu2021graph, choudhury2023tempo, shuai2023reconstructing, Liao_2024_CVPR}, they are still highly sensitive to occlusions, lighting variations, and the actor's appearance. Additionally, many of these assume a static camera setting, limiting capture space and reducing effectiveness in dynamic or outdoor environments. These constraints hinder the broader application of optical motion capture.

\subsubsection{Inertial-based MoCap} 
Inertial Measurement Units (IMUs) provide efficient and compact measurement for the orientation and acceleration of the human skeleton, facilitating accurate motion capture \cite{kamboj2024survey}. Commercial systems like Noitom \cite{noitom_website} and Xsens \cite{movella_website} achieve precise capture by embedding dense IMU arrays into tight-fitting bandages. However, the intrusive nature of these setups significantly restricts user comfort and accessibility. To address these challenges, recent research has focused on sparse-IMU solutions to balance functionality and user comfort. SIP reduced the IMU count to six, but limited to offline \cite{von2017sparse}. DIP introduced real-time performance using deep recurrent neural networks, but was constrained to local pose estimations \cite{huang2018deep}. TransPose improved global translations by incorporating foot-ground contact constraints \cite{yi2021transpose}. PIP \cite{PIPCVPR2022} and TIP \cite{jiang2022transformer} refine TransPose further: PIP ensures physical plausibility through physics-based optimization, while TIP enhances tracking accuracy using transformer architectures. PNP \cite{yi2024physical} improved the accuracy and robustness by compensating for fictitious forces in non-inertial frames. Furthermore, UIP \cite{armani2024ultra} combined IMUs with UWB to constrain drift and jitter, improving accuracy. IMUPoser \cite{mollyn2023imuposer} estimated pose by IMUs in devices like smartphones, smartwatches, and earbuds, eliminating the need for external sensors and further improving comfort and accessibility. However, its accuracy may decline when users lack or do not wear the required devices.

Despite these innovations, sparse-IMU systems still struggle with pose ambiguity, often misclassifying similar movements like sitting and standing, leading to errors in elbow and knee tracking. Moreover, they rely on tight straps for accuracy, compromising long-term comfort and wearability.

\subsubsection{Flexible Sensor-based MoCap}
Flexible sensors are emerging as promising solutions for long-term monitoring of human activities due to their insusceptibility to issues like lighting conditions, integration drift, or occlusions \cite{nag2017wearable, luo2022digital, li2024intelligent}. In addition, their inherent bio-compatibility, high stretchability, and lightweight design make them particularly suitable for monitoring of human physical status at extended duration, enhancing the overall user experience \cite{luo2021learning, luo2023technology}. Despite their comfort, flexible sensors may not yet match the precision of IMUs, as they are unable to directly measure the 3-axis orientation of bones, posing a less accurate full-body motion capture compared to inertial systems \cite{chen2023full}. However, their strength of directly measuring joint angles through skin deformation, renders them particularly accurate for joints with noticeable deformation, such as the elbow \cite{chen2023dispad, zuo2023self, jiaweisuda} and knee joints \cite{bergmann2013attachable, papi2018flexible, sheeja2023design}.

In general, their seamless integration into clothing ensures enhanced comfort and long-term wearability. Although flexible sensors have not yet surpassed IMUs in full-body motion capture accuracy, their exceptional capabilities in tracking knee and elbow joints offer a promising complement to inertial MoCap systems, potentially enhancing both accuracy and user satisfaction.

\subsubsection{Clothes-based MoCap}

Clothing has been indispensable to everyday life since humans first donned garments for warmth and protection during the Ice Age \cite{kittler2003molecular}. In the modern world, the always-on nature and portability of clothing make it an ideal medium for interacting with digital realities \cite{weiser1991computer}. Therefore, sensors are increasingly being embedded into garments to enable long-term MoCap. To obtain accurate data, mainstream approaches typically involve integrating either inertial sensors or flexible sensors at key joint locations within tight-fitting clothing. For instance, the TESLASUIT incorporates a dense array of IMUs (n=14) into its tight-fitting garment \cite{teslasuit_website}. On the other hand, Chen et al. demonstrate the use of six flexible sensors on an elbow pad to predict joint angles \cite{chen2023dispad}.

While tight-fitting clothing-based motion capture has shown promise, solutions for loose-fitting garments remain underexplored. LIP demonstrates robust accuracy by integrating a sparse set of IMUs (n=4) into a loose-fitting jacket \cite{zuo2024loose}, but its performance on the elbow joint is limited due to the sparse sensor placement. Other approaches, such as MoCaPose \cite{zhou2023mocapose} and SeamPose \cite{yu2024seampose}, incorporate capacitive sensors into loose garments for motion capture. However, these methods rely on a single sensor type, restricting their accuracy and adaptability. In general, as shown in Tab. \ref{tab:capacitive}, our approach achieves accurate and comforatble motion capture paradigm at a smooth 60 Hz by leveraging both IMUs and flex sensors in everyday clothing. This multimodal integration is intuitive, as garments naturally serve as an ideal interface for sensor fusion.



\begin{table}[h]
\caption{Comparison with other methods in loose-fitting clothing. Our FIP method demonstrates similar, or even superior, motion capture performance with higher frame rates.}
\label{tab:capacitive}
\begin{tabular}{ccccc}
\toprule
    Method & Sensor Types & FPS & Positional Error (cm) \\ \midrule
    MoCaPose & capacitive sensors    & 30* & 8.80 \\ 
    SeamPose & capacitive sensors    & 30* & 8.60 \\ 
    LIP      & IMUs                  & 30  & 9.99 ± 4.19 \\ 
    Ours     & IMUs and flex sensors & 60  & 8.06 ± 2.87 \\ \bottomrule
\multicolumn{4}{l} {\footnotesize *: According to the dataset.}
\end{tabular}
\end{table}

\subsection{Algorithms for Sensor Displacements}
Sensor displacements occur when sensors shift from their intended configuration, resulting in altered positioning that can challenge pre-trained pattern recognition models. To address this, researchers have employed transfer learning techniques \cite{wang2018deep, qin2019cross,kamboj2024fusion}, such as domain adaptation \cite{chang2020systematic, zhao2020local} and self-supervised learning \cite{jain2022collossl, deldari2022cocoa, haresamudram2022assessing}, to extract invariant features from displaced sensor data \cite{kunze2008dealing, wu2018orientation}. For IMUs, static pose calibrations, like A-pose and T-pose \cite{jiang2022transformer, yi2021transpose,noitom_website}, are commonly used to eliminate the initial relative rotation bias between the body and sensors, referred to as \textit{Primary Displacement}. On another front, LIP generates synthetic loose IMU data to address \textit{Real-time Displacement}. However, LIP relies on the assumption that the distributions of loose and tight IMU data are sufficiently similar, which may lead to the generated displacement distribution diverging from real-world conditions \cite{zuo2024loose}.

For flexible sensors, Zuo et al. proposed a self-adaptive algorithm that fine-tunes and aligns distributions to handle \textit{Real-time Displacement} \cite{zuo2023self}. DisPad leverages fuzzy entropy to identify the closest matching distribution to the current displacement, effectively managing both circular and lateral sensor shifts \cite{chen2023dispad}. Meanwhile, SuDA introduced a Sim2Real approach that aligns the support of source and target domains, rather than their distributions, achieving results comparable to supervised learning without the requirements for real labeled data \cite{jiaweisuda}.

To sum up, while these methods have been successful in addressing single-sensor displacement, handling displacement across multiple sensors remains a significant challenge. In this work, we propose targeted algorithms for both flex and inertial sensors, designed to account for their distinct displacement characteristics and ensure robust performance in everyday garments.

\subsection{Generative AI}

Generative AI has widespread applications across various fields, including image \cite{openai_dalle3, openai_dalle2}, speech \cite{cao2024survey}, and text generation \cite{openai_chatgpt}, with profound impact on the Human-Computer Interaction (HCI) community \cite{muller2022genaichi}. Variational Autoencoders (VAEs) \cite{kingma2013auto} and Diffusion Models (DMs) \cite{ho2020denoising}, both of which are designed to learn underlying data distributions and generate new samples from them. 

For VAEs, the model optimizes the Evidence Lower Bound (ELBO), formulated as:
\begin{equation}
\label{eq:sample}
\mathcal{L}(\xi, \phi; x) = \mathbb{E}_{q_\phi(z|x)}[\log p_\xi(x|z)] - D_{KL}(q_\phi(z|x) \parallel p(z)),
\end{equation}
where \(q_\phi(z|x)\) is the approximate posterior, \(p_\xi(x|z)\) is the likelihood of generating the data, and \(D_{KL}\) is the Kullback-Leibler divergence that regularizes the latent space to follow a prior distribution.

Diffusion Models, in contrast, work by modeling the reverse process of a gradual noise addition. The forward process adds Gaussian noise to the data over \(T\) timesteps, modeled as:
\begin{equation}
q(x_t|x_{t-1}) = \mathcal{N}(x_t; \sqrt{1 - \beta_t}x_{t-1}, \beta_t I)
\end{equation}
where \(\beta_t\) controls the noise schedule. The model learns the reverse process, recovering the original data by predicting the noise at each step:
\begin{equation}
\label{eq:sample}
p_\theta(x_{t-1}|x_t) = \mathcal{N}(x_{t-1}; \mu_\theta(x_t, t), \Sigma_\theta(x_t, t))
\end{equation}
By minimizing the difference between the added noise and the predicted noise, the model can generate data by starting from pure noise and iteratively denoising it.

Both VAEs and Diffusion Models are pivotal techniques in generative AI, driving advancements in generating high-quality, complex data across various domains. In our application, we seek to harness the capabilities of generative AI to create a diverse range of IMU displacements, which could enhance the robustness of pose estimation models.
\section{Design \& Prototyping of FIP}
\begin{figure}[ht]
\centering
\includegraphics[width=1.0\linewidth]{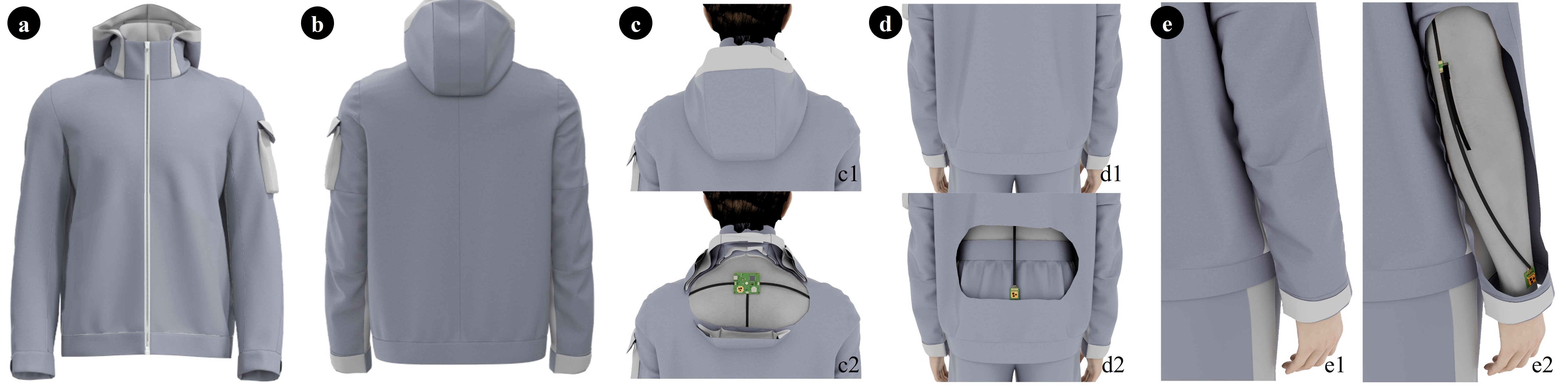}
\caption{Prototype of the garment integrated with flex sensors, IMUs, and circuit board: (a-b) Front and back views of the garment, (c) IMU and circuit board positioned on the back, (d) IMU located at the waist, (e) Flexible sensors placed on the elbows and IMUs attached to the forearms.}
\label{fig:hardware}
\end{figure}

As shown in Fig. \ref{fig:hardware}, our system prototype incorporates two flex sensors provided by Nitto \footnote{https://www.nitto.com/} and four Xsens \footnote{https://www.xsens.com/} MTI-3 IMUs into a jacket, endowing it with the ability for robust motion capture. The four IMUs are strategically placed on the left forearm, right forearm, back, and waist, while the two flex sensors are positioned at the left and right elbow. These sensors are connected with flexible wire and seamlessly integrated into the fabric through heat pressing, ensuring a non-invasive experience.

\subsection{Design}
Our design prioritizes both user comfort and data collection accuracy. Sensor placement and readout interfaces are meticulously selected based on joint movement characteristics for effective integration.

\paragraph{Sensor Placement} Drawing on previous research, we positioned four IMUs at key joints with three degrees of freedom: the left forearm, right forearm, back, and waist. Given that the elbow has only one degree of freedom and rigid sensors could compromise comfort, we incorporated flexible bending sensors at the elbows of both sleeves to enhance tracking accuracy.

\paragraph{Circuit Design} The sensor readout system comprises three components: the main control board, flex sensor connection board, and IMU connection board. The main control board, equipped with an ultra-low-power microcontroller, serves as the system's core, offering circuit protection, power management, and Bluetooth transmission. To minimize discomfort, the flex sensor connection board is designed in an "L" shape with dual-sided solder pads secured with adhesive to increase durability, while the IMU connection board has a rounded square shape with the solder pads placed on the back.


\begin{figure}[ht]
\centering
\includegraphics[width=1.0\linewidth]{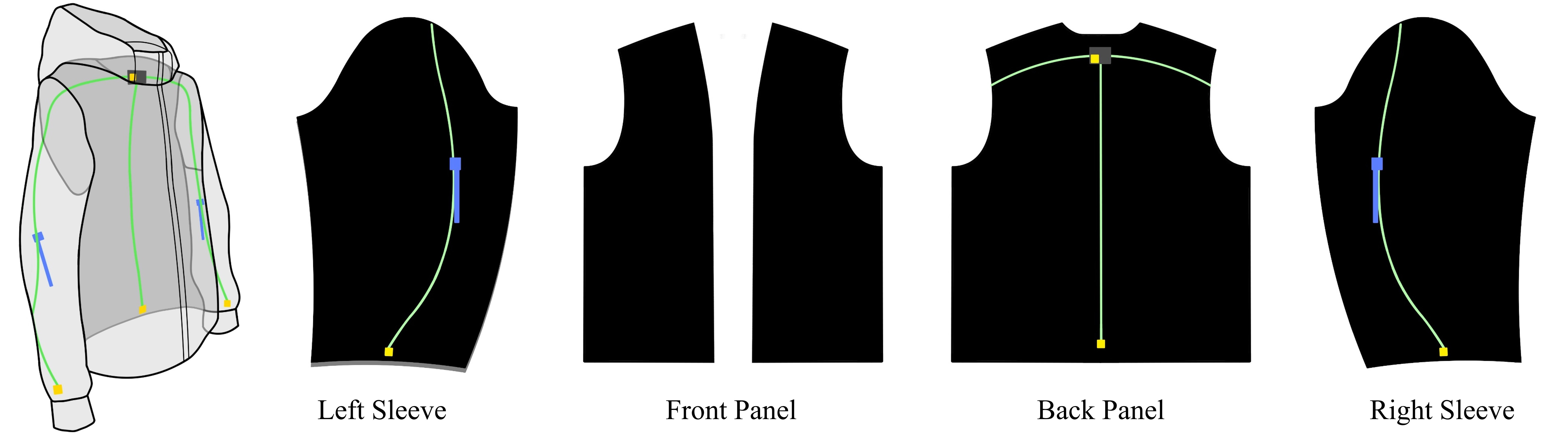}
\caption{Garment pattern design and sensor placement of our prototype. \textcolor[RGB]{91,126,254}{\textbf{Blue}}: Flex Sensor; \textcolor[RGB]{254,236,1}{\textbf{Yellow}}: IMU; \textcolor[RGB]{175,255,167}{\textbf{Green}}: Wire.}
\label{fig:garment_design}
\end{figure}


\paragraph{Garment Design} The design of the garment patterns for our prototype, including the placement of sensors on these patterns, is illustrated in Fig. \ref{fig:garment_design}. Through this pattern design, we defined the shapes of the fabric pieces required for garment construction. During the prototyping process, we will use this design to mark the placement of the sensors to ensure accurate integration.
\begin{figure*}[t]
\centering
\includegraphics[width=1.0\linewidth]{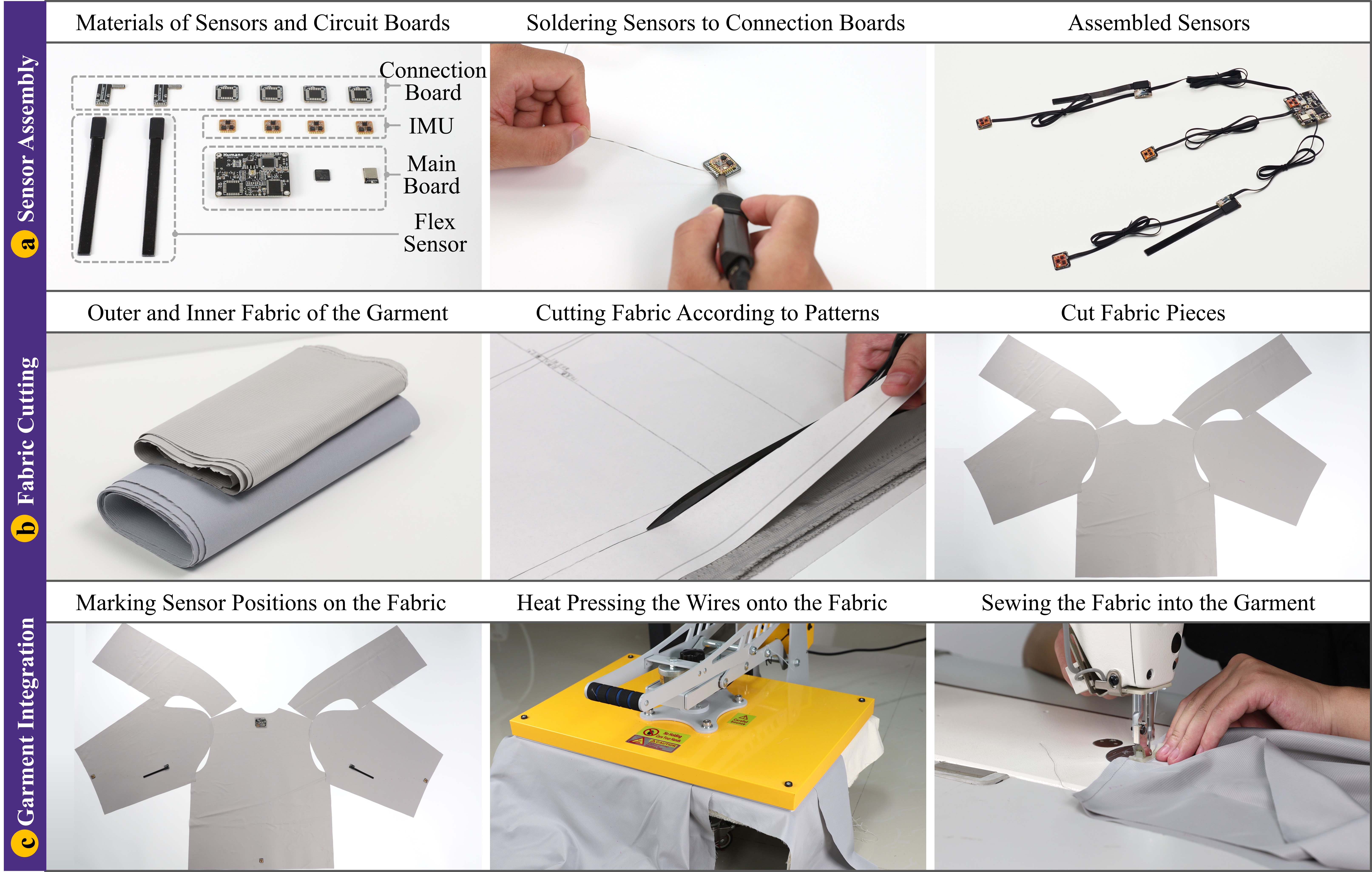}
\caption{The process of prototyping: (a) Assemble the sensors together by soldering; (b) Cut the fabric into pieces according to the patterns; (c) Integrate the assembled sensors into the fabric through heat pressing and sew the fabric into the garment.}
\label{fig:prototyping}
\end{figure*}

\subsection{Prototyping}
The prototype production consists of three stages: sensor assembly, fabric cutting, and garment integration (Fig. \ref{fig:prototyping}).

\paragraph{Sensor Assembly} First, the flex sensors and IMUs are soldered onto their respective connection boards. The main control board is then connected to the two flex sensors and four IMUs via two $I^2$C buses and four UART channels, completing the sensor assembly.

\paragraph{Fabric Cutting} Based on garment design, full-scale paper garment patterns are printed. These patterns are placed directly on the nylon fabric surface. The fabric is then cut according to these patterns, and the wiring and sensor placement are marked on the cut fabric pieces.

\paragraph{Garment Integration} The assembled sensors are integrated into the garment through heat pressing. First, the heat adhesive film is cut and heat-pressed onto the fabric. Next, the lining is sewn, and the adhesive film is bonded to the fabric to create wiring channels. Finally, the outer fabric is sewn to the lining, completing the double-layered smart garment.






\section{FIP for Sensor Displacements}

To address the sensor displacements in loose-fitting garments, FIP integrates three key components: 1) a Displacement Latent Diffusion Model (DLDM) for generating sufficient IMU \textit{Real-time Displacement} data to train a robust motion estimation model; 2) a Physics-informed Calibrator (PIC) to register flex sensor data in different \textit{Primary Displacements}; 3) a Pose Fusion Predictor (PFP) to fuse multi-modal sensor readings with elbow joint focus.

\begin{figure*}[h]
\centering
\includegraphics[width=1.0\linewidth]{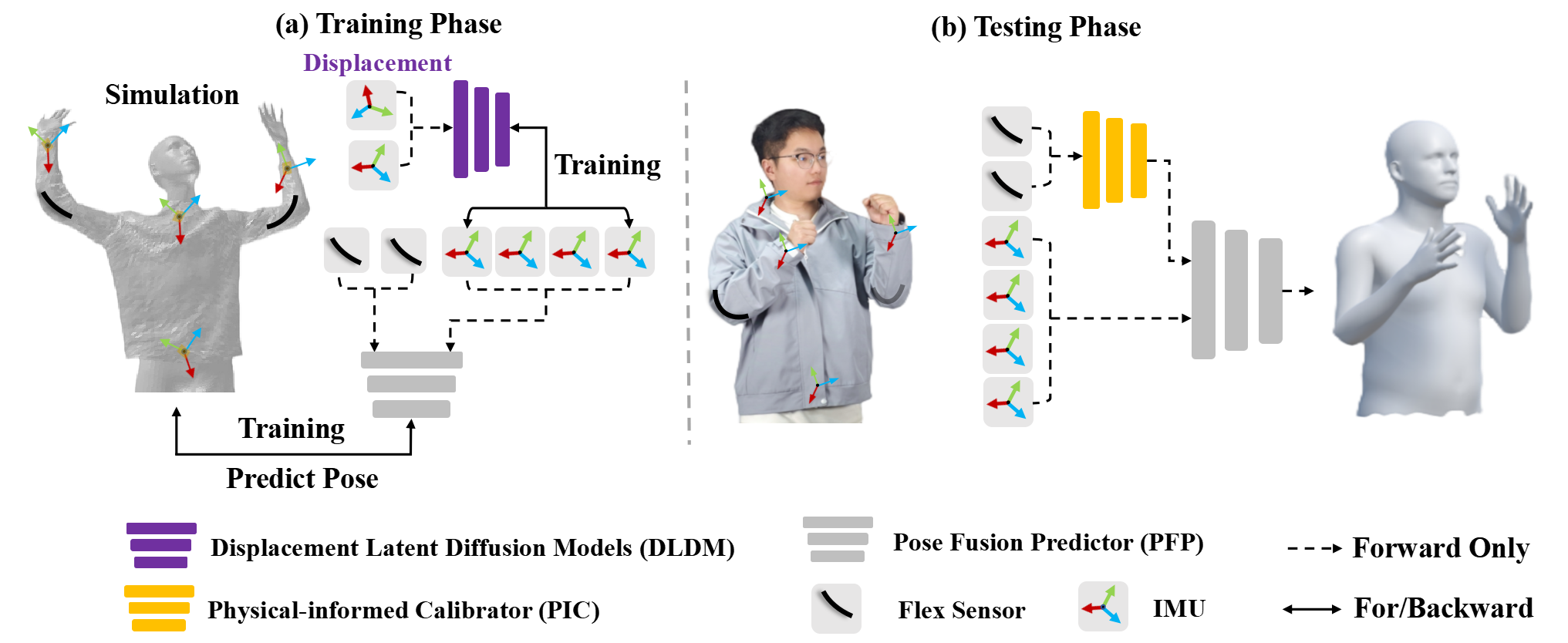}
\caption{Pipeline Overview. Left: For data preparation, we first utilize a simulation body-fabric model to synthesize the IMU \textit{Real-time Displacement}. Then, for training a robust pose predictor, we train a Displacement Latent Diffusion Model (DLDM) to generate enough diverse data that covers real-world distribution. At last, we train a Pose Fusion Predictor leveraging simulated flex sensor data and generated IMU data, with the supervision of SMPL Pose \cite{loper2023smpl}. Right: In our testing phase, flex sensor readings will be firstly input to our Physical-informed Calibrator to address the \textit{Primary Displacement}, which will be input to the pre-trained Pose Fusion Predictor with IMU data.}
\label{fig:framework}
\end{figure*}

Fig. \ref{fig:framework} illustrates the overview of our pipeline. Overall, our objective is to estimate human body posture using readings from flex sensors and IMUs integrated into a loose-fitting daily garment. The input to our model includes measurements from 4 IMUs ($\boldsymbol{\mathbf{IMU} \in \mathbb{R}^{4 \times 9}}$) and 2 flex sensors ($\boldsymbol{\mathbf{Flex} \in \mathbb{R}^{2 \times 1}}$). The output of the model is the upper body postures represented by the 3D rotation angles of 10 joints ($\theta \in \mathbb{R}^{10 \times 3}$).



\paragraph{Training Phase} Firstly, as capturing IMU \textit{Real-time Displacement} in the real world is very challenging, we utilize a simulation fabric-body model to synthesize displacement data, denoted as $\mathbf{IMU}_{sim}$. Then, for training a robust pose predictor, we train a Displacement Latent Diffusion Model (DLDM) to generate enough diverse data that covers real-world distribution. In addition, we simulate the flex sensor readings $\mathbf{Flex}_{sim}$ by calculating the angle between the forearm and upper arm skeleton. Flex sensor data $\mathbf{Flex}_{sim}$ and generated IMU data $DLDM(\mathbf{IMU}_{sim})$ will be input to train a Pose Fusion Predictor (PFP), with the supervision from simulated SMPL Pose. The training phase can be denoted as: 

\begin{equation}
\begin{aligned}
\boldsymbol{\theta} = PFP(DLDM(\mathbf{IMU}_{sim}), \mathbf{Flex}_{sim})
\end{aligned}
\end{equation}

\paragraph{Testing Phase} Physical-informed Calibrator (PIC) is only employed in the real-world testing phase, to correct the flex sensor readings into real motion ranges. Then, the flex and IMU sensor readings $\mathbf{IMU}$ and calibrated $\mathbf{Flex}$ will input to PFP for the estimation of pose $\theta$, denoted as:
\begin{equation}
\begin{aligned}
\boldsymbol{\theta} = PFP(\mathbf{IMU}, PIC(\mathbf{Flex}))
\end{aligned}
\end{equation}


\subsection{Inertial Sensors: Displacement Latent Diffusion Model}


\textit{\textbf{Challenge:} Real-time displacement occurs as sensors shift and vibrate due to loose-fitting clothing. However, developing a machine learning model for real-time displacement presents significant challenges: 1) capturing real-time displacement data in real-world scenarios is highly difficult, and 2) the variability in user poses, body shapes, clothing fits, and random movements introduces substantial randomness, complicating the accurate simulation of real-time displacement.
}

\begin{figure*}[ht]
\centering
\includegraphics[width=1.0\linewidth]{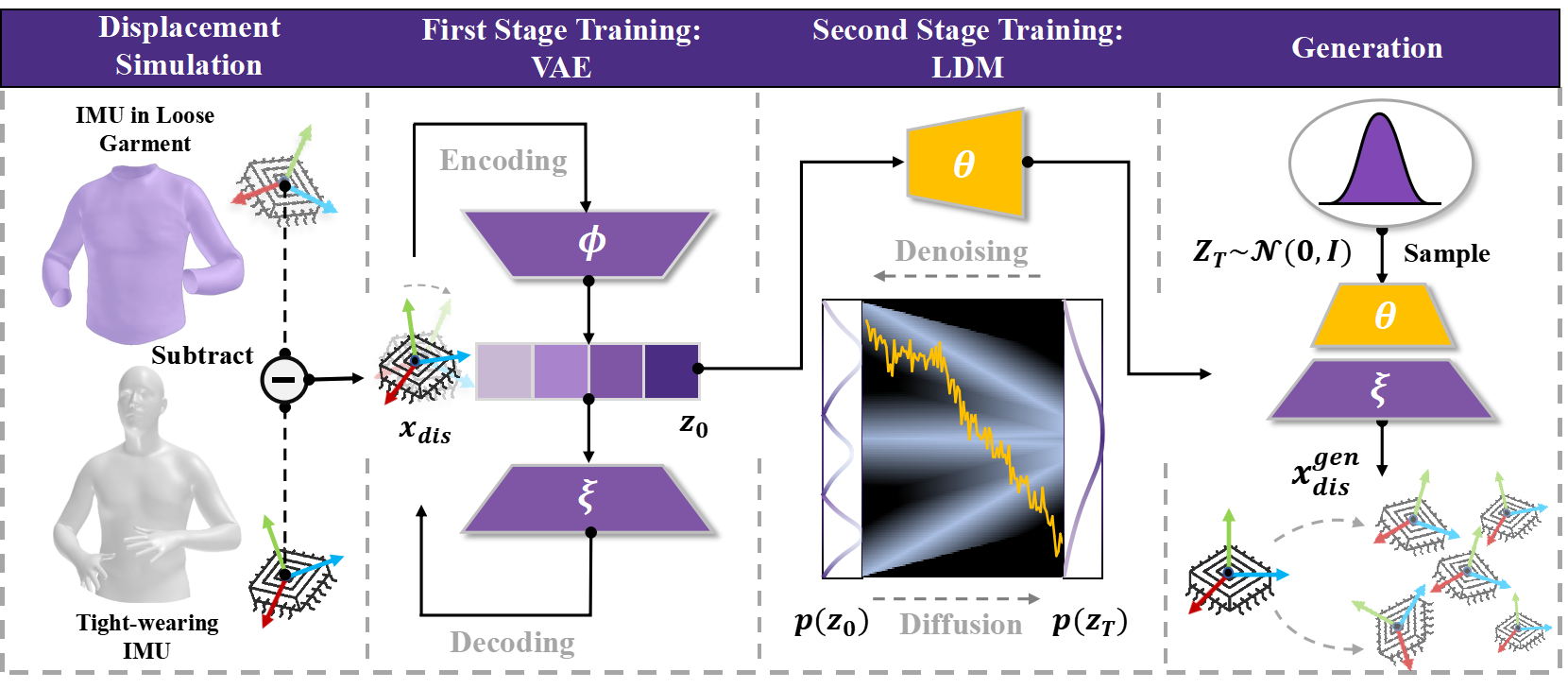}
\caption{Overview of Displacement Latent Diffusion Model (DLDM). (a) Firstly, the \textit{Real-time Displacement} $x_{dis}$ could be captured through subtracting paired Loose and Tight IMU simulation data; (b) After obtaining displacement $x_{dis}$, we trained a Variational Autoencoder (VAE) to encode an expressive displacement latent $z_0$  with respect to the encoder and decoder parameters $\phi$ and $\xi$; (c) Then, we freeze the VAE’s encoder and decoder networks and train Latent Diffusion Models (LDM) on the sampled latent $z_0$, which gradually become normal gaussian distribution $z_T \sim \mathcal N(0,I)$ through noise diffusion process. (d) Finally, with the expressive LDMs, we can define a hierarchical generative model to generate various displacements $x^{gen}_{dis}$ with LDM's denoising network $\theta$ and VAE's Decoder $\xi$ from Gaussian noise $z_T$. }
\label{fig:DLDM}
\end{figure*}

Fig. \ref{fig:DLDM} illustrates the overview our Displacement Latent Diffusion Model (DLDM). Our approach leverages the powerful capabilities of deep generative models to learn a representative latent space of IMU displacement $z$ from displacement $x_{dis}$. From this latent space $z$, we can generate diverse and realistic IMU displacement signals $x^{gen}_{dis}$, enabling the training of a robust pose estimation model.
To achieve this, we first establish a simulation framework to retrieve simulated IMU displacement data $x_{dis}$. Then, we train the DLDM via a dual-stage process that hierarchically optimizes $x_{dis}$ into a standard Gaussian distribution $z$. In the first stage, we train the model as a regular Variational Autoencoder (VAE) \cite{kingma2013auto} with standard Gaussian priors. Subsequently, we refine the latent encodings by training the latent Denoising Diffusion Models (DDMs) \cite{ho2020denoising}.

\subsubsection{Displacement Simulation}
As collecting the IMU displacement data in the real world is very challenging, we propose a simulation framework to obtain IMU \textit{Realtime Displacement} data.
Firstly, we simulate tight-wearing IMU signals by directly placing the virtual IMUs on the human body mesh obtained from the AMASS dataset \cite{mahmood2019amass}, and then utilize TailorNet \cite{patel2020tailornet} to simulate topologically consistent clothing given SMPL pose and body shape to obtain loose IMU data as described in \cite{zuo2024loose}. Finally, we subtract the simulation IMU in the loose garment and tight-wearing IMU data to obtain the real-time displacement $x_{dis}$.

\subsubsection{First Stage Training}
In the first stage of training, FIP is trained by optimizing the encoder and decoder parameters $\phi$ and $\xi$ to learn an expressive IMU displacement latent $z_{0}$ by minimizing:
\begin{equation}
\mathcal{L}(\xi, \phi; x_{dis}) = \mathbb{E}_{q_\phi(z_0|x_{dis})}[\log p_\xi(x|z_0)] - \lambda_{KL} D_{KL}(q_\phi(z_0|x) \parallel p(z_0)),
\end{equation}

Here, the IMU displacement latent $z_{0}$ is sampled from the encoded posterior distribution $q_{{\phi}}(z_{0}|x_{dis})$, whose means and variances are predicted via VAE's encoder.  Furthermore, $p_{\xi}(x|z_{0})$ denotes the decoder, parametrized as a Laplace distribution with predicted means and fixed unit scale parameter. $\lambda_{KL}$ are hyperparameters balancing reconstruction accuracy and Kullback-Leibler regularization. 


\subsubsection{Second Stage Training}
In principle, we could use the VAE’s priors to sample encodings and generate various displacements. However, we find that simple displacement latent from VAE does not accurately match the real-world displacement distribution from the training data, leading to poor samples (\textit{i.e.}, the prior hole problem) \cite{vahdat2021score}. This limitation motivates the use of highly expressive latent diffusion models (LDMs). Specifically, in the second stage, we freeze the VAE’s encoder and decoder networks and add Gaussian noise on the encodings \(z_0\) sampled from \(q_\phi(z_0|x)\) gradually until \(z_0\) becomes normal gaussian distribution $p(z_T) \sim \mathcal N(0,I)$. Then,  we could train a denoising diffusion model $\theta$ to reverse the diffusion process, which translates the latent $p( z_T)$ back to the original IMU displacement latent space.




\subsubsection{Generation}

With the expressive LDMs, we can formally define a hierarchical generative model  $p_{\xi,\theta}(x, z_0) = p_{\xi}(x | z_0) p_{\theta}(z_0)$ denotes the distribution of the displacement latent, and $p_\xi(x|z_0)$ is FIP’s Decoder. We can hierarchically sample from latent and generate diverse IMU displacements $x_{dis}^{gen}$ with the VAE decoder. 

With the generated diverse IMU displacements data $x_{dis}^{gen}$, we could train our robust pose estimation model, Pose Fusion Predictor. We will discuss the data and model architecture of VAE and LDMs in detail in Sec. \ref{sec:Expriments}.


\subsection{Flex Sensors: Physics-informed Calibrator}

\begin{figure}[ht]
\centering
\includegraphics[width=1.0\linewidth]{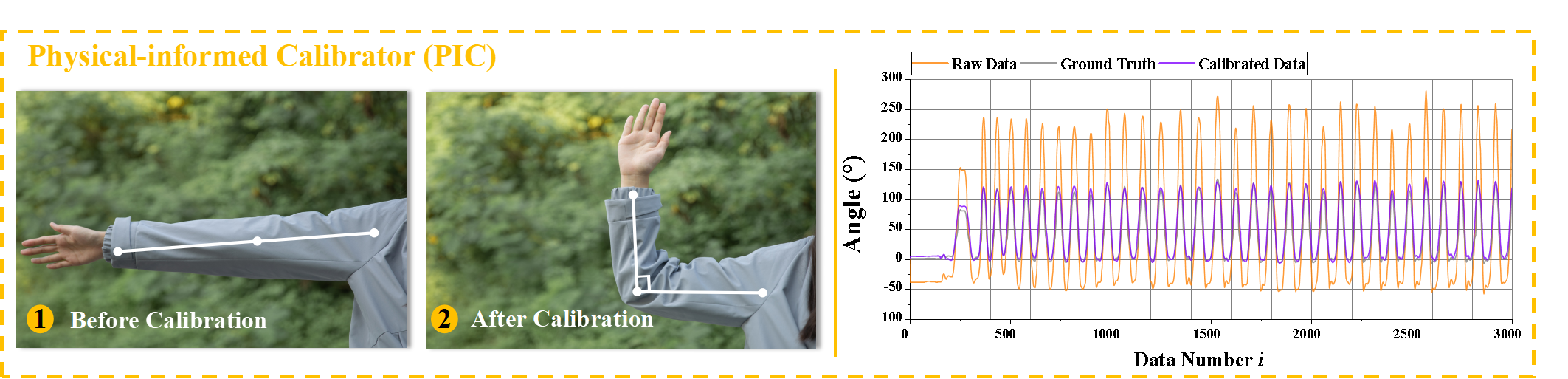}
\caption{The process and effect of Physics-informed Calibrator. Through one elbow flexion calibration, our PIC could rectify flex sensors readings effectively.}
\label{fig:pic}
\end{figure}

\textit{\textbf{Challenge:} Primary displacement during clothing donning alters sensor placement due to significant body-clothing contact, causing substantial data distribution discrepancies. This challenges models trained on synthetic displacement data to capture real-world displacement features. Hence, efficient calibration of Primary Displacement is crucial for accurate measurements with flex sensors.}

To address this, we introduce a Physics-informed Calibrator (PIC), which calibrates \textit{Primary Displacement} through a single elbow flexion. The PIC module is designed and implemented based on our observation that, despite significant shifts in sensor data range due to \textit{Primary Displacement}, the displaced sensor data are always aligned with the overall movement patterns. Therefore, by performing a full vertical elbow bend, we can recalibrate the sensor data to match the actual motion angles by realigning the data range, ensuring the measurements remain physically accurate.
In general, our PIC module can be written as:

\begin{equation}
\begin{aligned}
\theta^{C}=PIC(\theta)=\frac{\theta-\theta_{min}}{\theta_{max}-\theta_{min}} \times(\theta^{C}_{max}-\theta^{C}_{min}) + \theta^{C}_{min}
\end{aligned}
\label{eq:PIC}
\end{equation}
which $\theta^{C}$ represents the sensor readings after PIC, and the minimum and maximum range for calibrating data are denoted as $\theta^{C}_{min}$ and $\theta^{C}_{max}$. In our setup, the user performs a full vertical elbow bend, so we define $\theta^{C}_{min}=0$ and $\theta^{C}_{max}=90$. Similarly, $\theta^{raw}$ refers to the raw sensor data, and its minimum and maximum range $\theta_{min}$, and $\theta_{max}$ is collected in our calibration process, respectively. According to the Eq.~\ref{eq:PIC}, we could scale the raw sensor data $\theta\in(\theta_{min},\theta_{max})$ to the corrected motion angle data $\theta^{C}\in(\theta^{C}_{min},\theta^{C}_{max})$, ensuring physical accuracy. 

\begin{equation}
\begin{aligned}
\frac{\theta^{C}-\theta^{C}_{min}}{\theta^{C}_{max}-\theta^{C}_{min}}=\frac{\theta-\theta_{min}}{\theta_{max}-\theta_{min}}
\end{aligned}
\end{equation}


\subsection{Pose Fusion Predictor (PFP)}
\begin{figure*}[ht]
\centering
\includegraphics[width=1.0\linewidth]{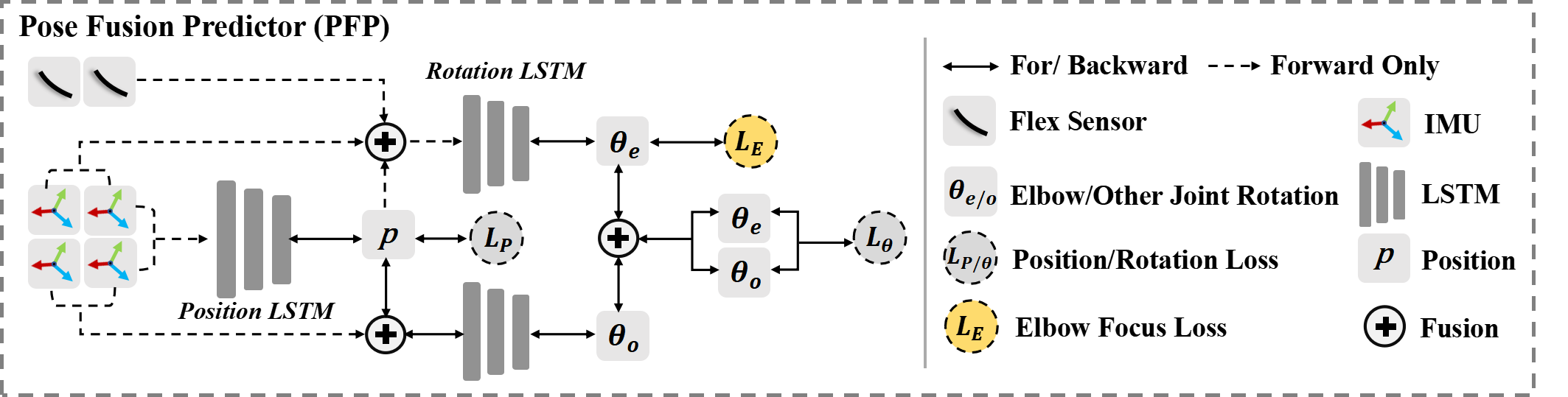}
\caption{Structure of Pose Fusion Predictor. First, the Position LSTM uses four IMU readings to predict joint positions \(\mathbf{\textit{p}} \in \mathbb{R}^{11 \times 3}\). The rotation prediction is then split into two stages. One Rotation LSTM combines \(\mathbf{Flex}\), \(\mathbf{IMU}\), and \(\mathbf{\textit{p}}\) to predict elbow joint rotations \(\boldsymbol{\theta_e} \in \mathbb{R}^{2 \times 3}\), while the other combines \(\mathbf{p}\) and \(\mathbf{IMU}\) to predict the remaining joint rotations \(\boldsymbol{\theta_o} \in \mathbb{R}^{8 \times 3}\). Finally, \(\boldsymbol{\theta_e}\) and \(\boldsymbol{\theta_o}\) are concatenated to produce the full joint rotation output \(\boldsymbol{\theta}\). PFP is trained by Position Loss $  \mathcal{L}_p$ and Rotation Loss $\mathcal{L}_\theta$, with an additional Elbow Focus Loss  $ \mathcal{L}_E$ to improve the predictions of elbow joints.
}
\label{fig:pose_predictor}
\end{figure*}

The structure of the Pose Fusion Predictor is depicted in Fig. \ref{fig:pose_predictor}. The goal of the PFP is to estimate the 3D joint rotations \(\boldsymbol{\theta}\in \mathbb{R}^{10 \times 3}\) using the flex and inertial sensor readings \(\mathbf{Flex}\) and \(\mathbf{IMU}\). It is composed of 3 long short-term memory (LSTM) layers.

Firstly, the Position LSTM utilizes four Inertial sensor readings $\mathbf{IMU}$ to predict joint positions $\mathbf{\textit{p}} \in \mathbb{R}^{11 \times 3}$. Then we decompose our $\theta$ prediction into two stages. The first Rotation LSTM combines \(\mathbf{Flex}\),  \(\mathbf{IMU}\), and \(\mathbf{\textit{p}}\) to predict the two elbow joint rotations \(\boldsymbol{\theta_e} \in \mathbb{R}^{2 \times 3}\), whereas another Rotation LSTM combines $\mathbf{\textit{p}}$ and \(\mathbf{IMU}\) to predict other joint rotations \(\boldsymbol{\theta_o} \in \mathbb{R}^{8 \times 3}\). $\boldsymbol{\theta_e}$ is then concatenated with $\boldsymbol{\theta_o}$ to form the final joint rotation output \(\boldsymbol{\theta}\).

To improve the predictions of elbow joints, which obtain a large degree of freedom and movement range, we include an additional Elbow Focus Loss $\mathcal{L}_{E}$. This loss encourages the attention to the elbow joint by computing the elbow bending angle through forward kinematics when training the first Rotation LSTM. 

To train our PFP, the total loss $\mathcal{L}$ is defined as a weighted sum of three components: $\mathcal{L}_{p}$ corresponds to the prediction of joint positions \( \textit{p} \); $\mathcal{L}_{\theta}$ corresponds to the prediction of joint rotations $\theta$; $\mathcal{L}_{E}$ ensures the model pays special attention to accurate elbow rotation prediction. All losses are Mean Squared Error (MSE) between ground truth and prediction.
The full loss function can be expressed as:

\begin{equation}
\begin{aligned}
\mathcal{L} = \lambda_1 \mathcal{L}_{p} + \lambda_2 \mathcal{L}_{\theta} + \lambda_3 \mathcal{L}_{E}
\end{aligned}
\end{equation}

where $\lambda_1, \lambda_2, \lambda_3$ are the weighting factors that control the influence of each component. By minimizing this total loss, our PFP is trained to enhance overall pose estimation accuracy, with a particular focus on elbow prediction.

\section{Experiments}
\label{sec:Expriments}
In this section, we first describe the dataset, metrics, and training details used in our approach (Sec. \ref{sec:dataset}). We then present the results of our method compared to SOTA real-time posture estimation methods (Sec. \ref{sec:results}), followed by an ablation study to analyze the impact of proposed techniques (Sec. \ref{sec:ablation}). Finally, we demonstrate the real-time motion capture interface of our approach (Sec. \ref{sec:realtime}).

\subsection{Implementation}
\label{sec:dataset}

\subsubsection{Dataset} 
\begin{figure*}[ht]
\centering
\includegraphics[width=1.0\linewidth]{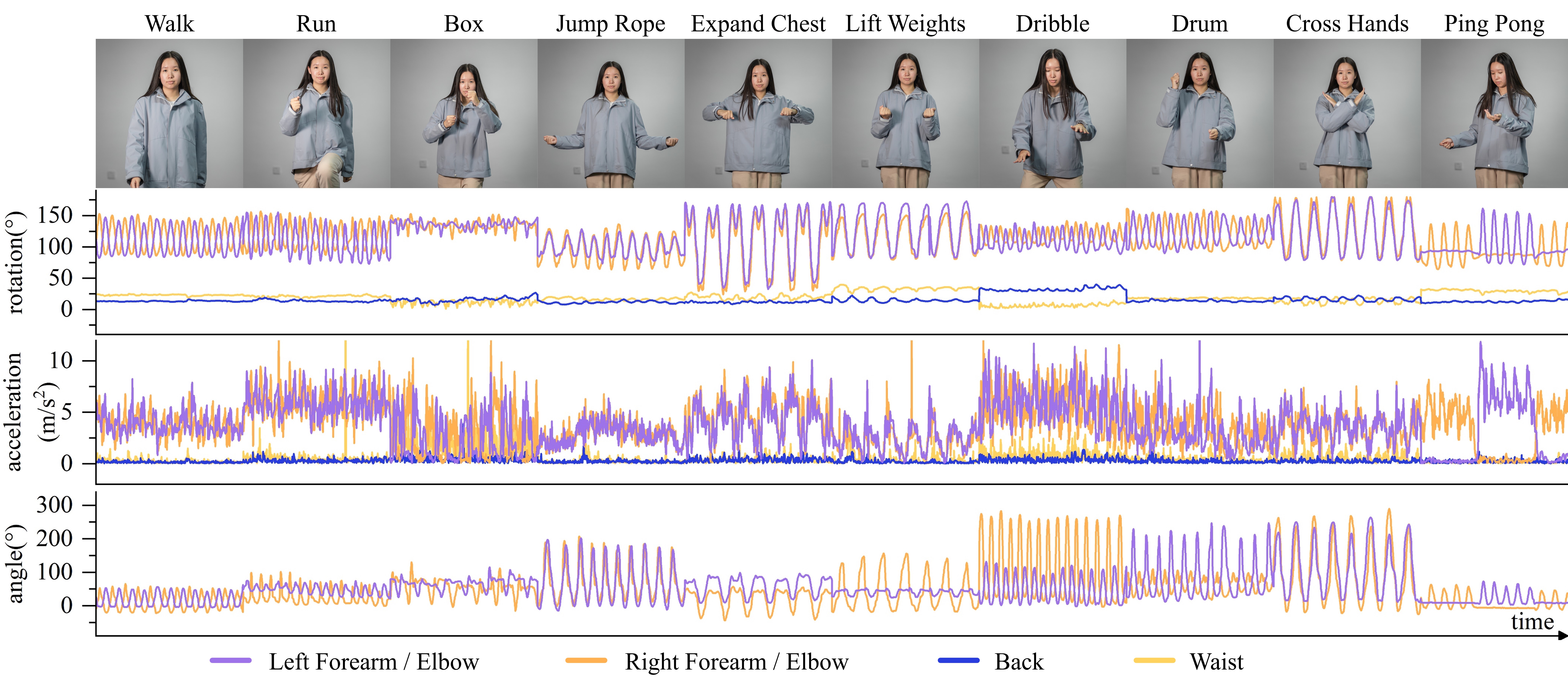}
\caption{Flex sensor and IMU signals. We present the sensor readings from both the flex sensors and IMUs while the user performs various actions, including walking, running, boxing, jumping rope, expanding the chest, lifting weights, dribbling, drumming, crossing hands, and playing ping pong.}
\label{fig:motion}
\end{figure*}

Our dataset consists of two parts: the Simulation Dataset $D_{sim}$, $D_{sim}^{dis}$ and the Real Dataset $D_{real}$. The $D_{sim}$ and $D_{sim}^{dis}$ are employed for training our models, whereas the real data $D_{real}$ is utilized for testing purposes. 
\begin{itemize}
    \item \textbf{$D_{sim}$}: We simulated tight-wearing IMU signals by directly placing the virtual IMUs on the human body mesh obtained from the AMASS dataset \cite{mahmood2019amass}, using the method described above. Notably, we adapted the joint and mesh vertex settings to match our upper body IMU setup (left forearm, right forearm, back, and waist).
    
    \item \textbf{$D_{sim}^{dis}$}: Collecting the IMU displacement data in the real world is challenging as we could not directly capture the paired real-time tight-wearing and loose-wearing IMU data. Hence, we firstly followed the methodology developed for loose IMU sensors based on $D_{sim}$ as described in LIP \cite{zuo2024loose}, which utilized TailorNet to simulate topologically consistent clothing given SMPL pose and body shape. Then, we subtracted the simulation loose IMU data and the aforementioned tight-wearing IMU data to obtain the real-time disturbance.
    
    \item \textbf{$D_{real}$}: We recruited a total of ten individuals with varying body shapes and collected a real-world dataset for evaluation using our flex-and-IMU fusion garment. All participants were informed of the experiment's purpose and signed the consent agreement for participation. During data collection, participants were asked to perform ten predefined actions, including walking, running, boxing, and more. Each action lasted for one minute. We present the sensor readings while the user performs various actions in Fig. \ref{fig:motion}. For the ground truth pose, we used the Perception Neuron 3 system \footnote{https://neuronmocap.com/pages/perception-neuron-3/} to capture upper-body poses using 11 tight-wear IMUs. Overall, we collected 371,122 frames of 60 fps data, with a total duration of about 2 hours. More information can be found in the supplementary materials.
\end{itemize}

\subsubsection{Evaluation Metrics} Following TransPose \cite{yi2021transpose}, we measure the accuracy of pose estimation using the following four metrics: 
\begin{itemize}
    \item Angular Error ($^{\circ}$), which measures the mean rotation error of ten upper body joints in degrees; 
    \item Positional Error (cm), which measures the mean Euclidean distance error of 12 upper body joint endpoints in centimeters. Note that in practice, we only calculate the positional error of eleven joint endpoints as the position of the pelvis joint is kept at [0, 0, 0]. 
    \item Elbow Angular Error ($^{\circ}$), which measures the mean rotation error of two elbow joints in degrees.
    \item Jitter (m/$s^3$), which measures the average jerk of 12 upper body joints.
\end{itemize}
 
\subsubsection{Training Details} All our experiments run on a computer with an Intel(R) Core(TM) i7-13700K CPU and an NVIDIA RTX 4080 GPU. The model is implemented using PyTorch 2.4.0 with CUDA 11.8. Please refer to the supplementary material for the training details and PFP network structure.
\begin{itemize}
    \item  Training the Displacement Latent Diffusion Model. VAE and Diffusion Models were both trained with $D_{sim}^{dis}$, utilizing a batch size of 512. We set $\lambda_{KL} = 1e^{-8}$ to prevent over-regularization and enhance the fidelity of the generated results. The Adam optimizer was used with a learning rate of $lr = 1e^{-3}$ during training.
    \item  Training Pose Fusion Predictor. The pose estimation network was trained with $D_{sim}^{dis}$ and $D_{sim}$, utilizing a batch size of 512. We set $\lambda_{1} = 4$, $\lambda_{2} = 1$, and $\lambda_{3} = 0.1$. The Adam optimizer was used with a learning rate of $lr = 5e^{-4}$ and weight decay = $1e^{-5}$ during training. 
\end{itemize}

\begin{figure*}[t]
\centering
\includegraphics[width=1.0\linewidth]{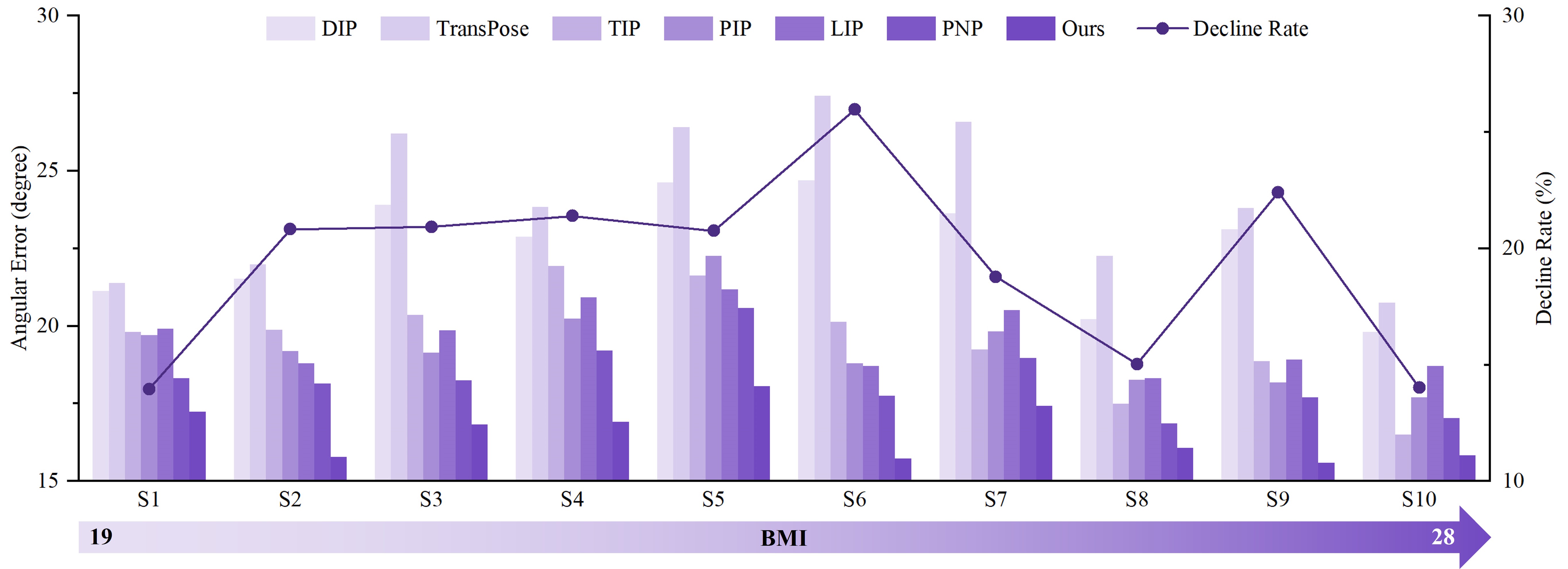}
\caption{Experimental results of Angular Error (unit: degree) on ten participants. The S1 indicates the subject with ID=1. The Decline Rate reflects the rate at which the error decreases in FIP.}
\label{fig:angle_err}
\end{figure*}
\begin{figure*}[t]
\centering
\includegraphics[width=1.0\linewidth]{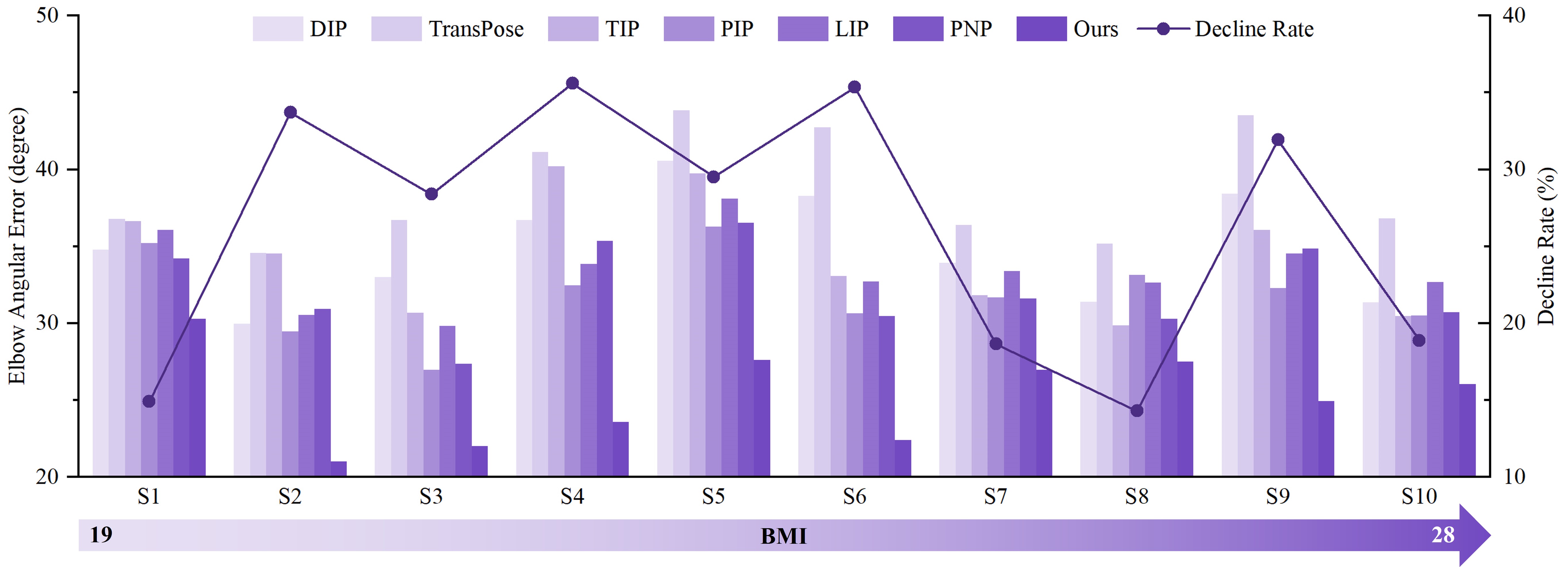}
\caption{Experimental results of Elbow Angular Error (unit: degree) on ten participants. The S1 indicates the subject with ID=1. The Decline Rate reflects the rate at which the error decreases in FIP.}
\label{fig:elbow_err}
\end{figure*}
\begin{figure*}[t]
\centering
\includegraphics[width=1.0\linewidth]{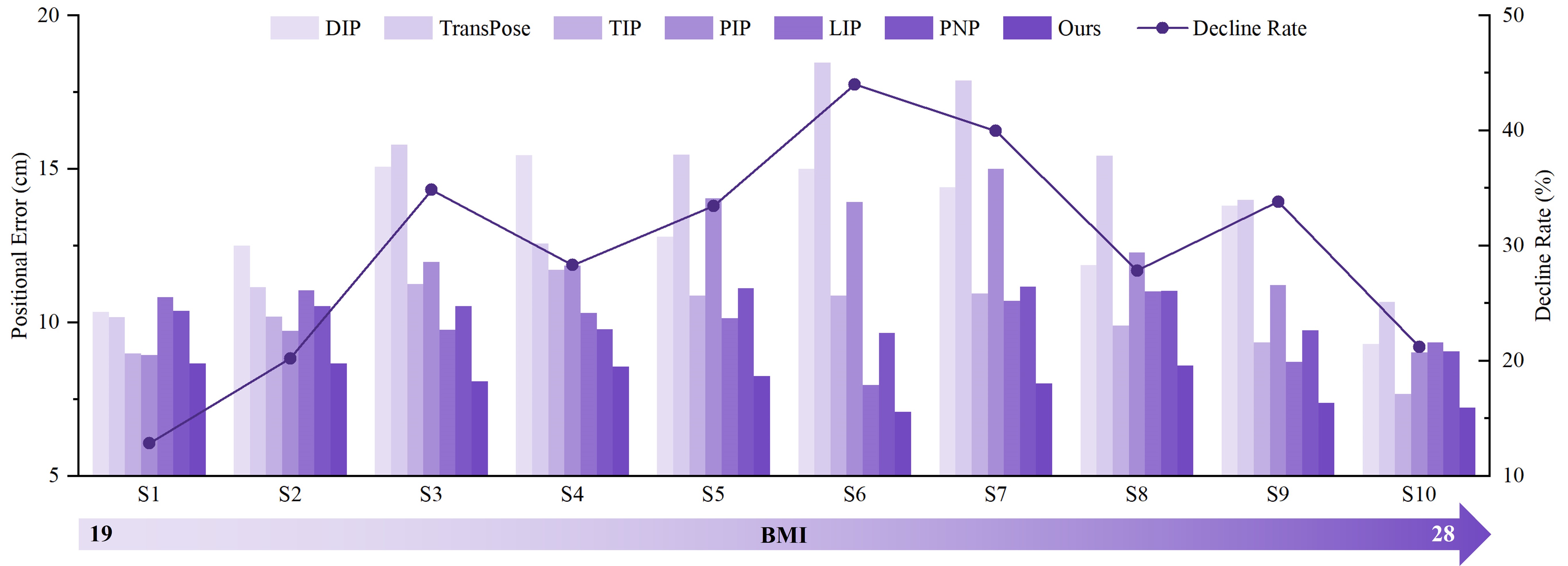}
\caption{Experimental results of Positional Error (unit: cm) on ten participants. The S1 indicates the subject with ID=1. The Decline Rate reflects the rate at which the error decreases in FIP.}
\label{fig:pos_err}
\end{figure*}

\begin{table*}[h]
\caption{Comparison with SOTA real-time \textit{Inertial-based MoCap} methods. Our approach significantly reduces elbow tracking errors and achieves SOTA performance in overall joint motion capture, surpassing all competing approaches.}
\label{tab:compare}
\begin{tabular}{cccccc}
\toprule
System    & Angular Error ($^{\circ}$)         & Elbow Angular Error ($^{\circ}$)    & Position Error (cm)       & Jitter (m/$s^3$)       \\ \midrule
DIP       & 22.53 ± 5.26          & 34.90 ± 13.25          & 13.02 ± 5.17         & 0.34          \\
TransPose & 23.98 ± 5.57          & 38.84 ± 12.11          & 14.03 ± 6.05         & 0.13          \\
TIP       & 19.61 ± 6.29          & 34.49 ± 14.00          & 10.16 ± 4.60         & 2.00          \\
PIP       & 19.29 ± 6.04          & 31.95 ± 15.77          & 11.72 ± 5.27         & 0.37          \\
LIP       & 19.58 ± 5.03          & 33.50 ± 12.48          & 9.99 ± 4.19          & 0.17          \\
PNP       & 18.29 ± 4.17          & 32.37 ± 12.44          & 10.29 ± 2.91         & 0.36          \\
Ours      & \textbf{16.54 ± 3.76} & \textbf{25.27 ± 11.15} & \textbf{8.06 ± 2.87} & \textbf{0.05} \\ \bottomrule
\end{tabular}
\end{table*}

\subsection{Results} 
\label{sec:results}


Our model predicts upper body poses with an angular error of 16.54$^{\circ}$ (a 19.5 \% improvement), an elbow angular error of 25.27$^{\circ}$ (a 26.4 \% improvement), and a position error of 8.06 cm (a 30.1 \% improvement) compared with SOTA real-time \textit{Inertial-based MoCap} methods, including DIP \cite{huang2018deep}, Transpose \cite{yi2021transpose}, TIP \cite{jiang2022transformer}, PIP \cite{yi2022physical}, LIP \cite{zuo2024loose}, and PNP \cite{yi2024physical}. Since most of these SOTA methods are designed for tight-wear IMUs with different sensor numbers and installation positions, we followed the official code provided by the authors and retrained the models using $D_{sim}$ and reported the angular and positional errors on $D_{real}$.  For LIP, we adopted the released model without any change.

As illustrated in Fig. \labelcref{fig:angle_err,fig:elbow_err,fig:pos_err} and Tab. \ref{tab:compare}, our method surpasses all SOTA motion capture techniques, including those using tight-wearing IMU methods (e.g., PIP, PNP) and previous clothing-based IMU approaches (e.g., LIP). This demonstrates a significant advantage in robustness across a diverse range of body shapes and motions.

\begin{figure*}[ht]
\centering
\includegraphics[width=1.0\linewidth]{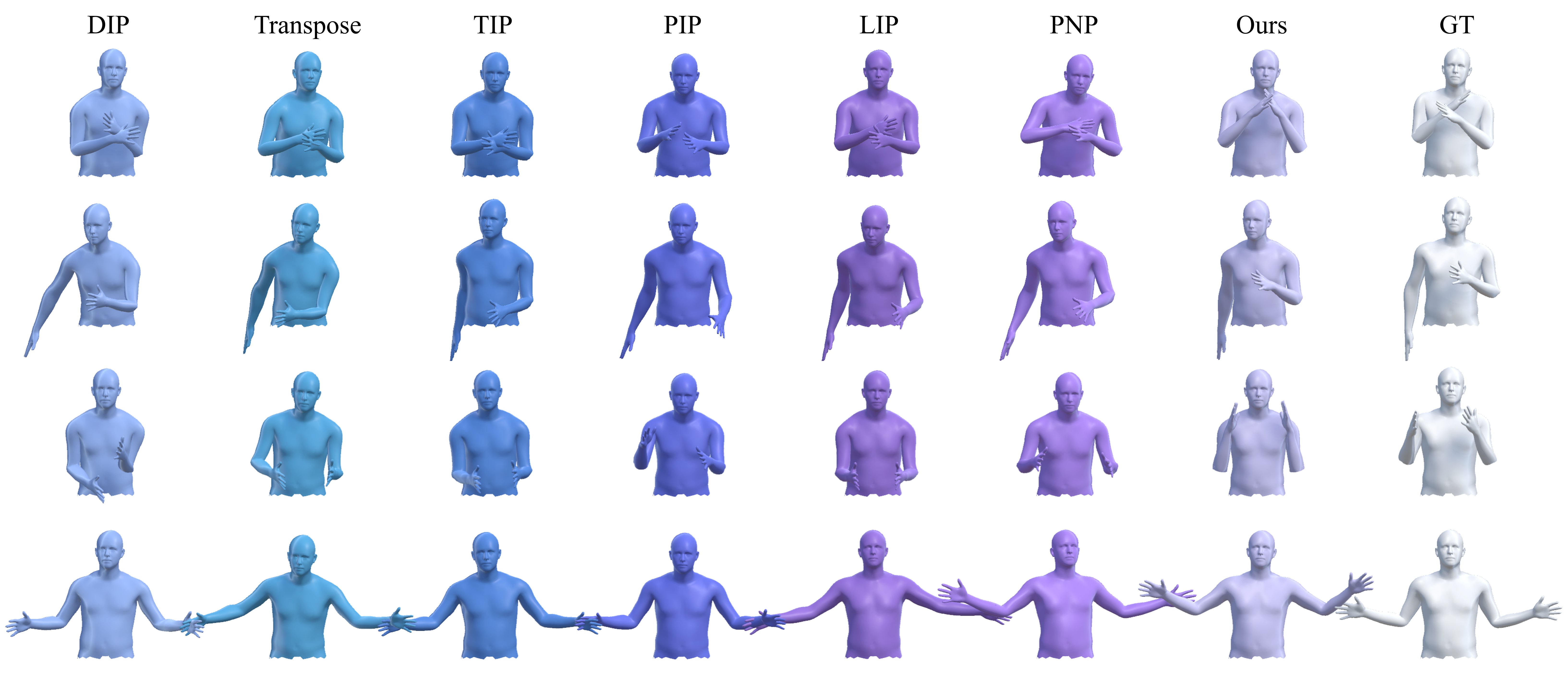}
\caption{Qualitative results: our approach outperforms all SOTA methods in motion capture with a clear advantage in elbow joint tracking.}
\label{fig:visualization}
\end{figure*}

Notably, as shown in Fig. \ref{fig:elbow_err} and Fig. \ref{fig:visualization}, our method can substantially reduce the elbow angular error, which is crucial for various real-world applications. This underscores the benefits of integrating IMUs with sensors in our approach.

\subsection{Ablation Study}
\label{sec:ablation}
\begin{figure}[ht]
\centering
\includegraphics[width=1.0\linewidth]{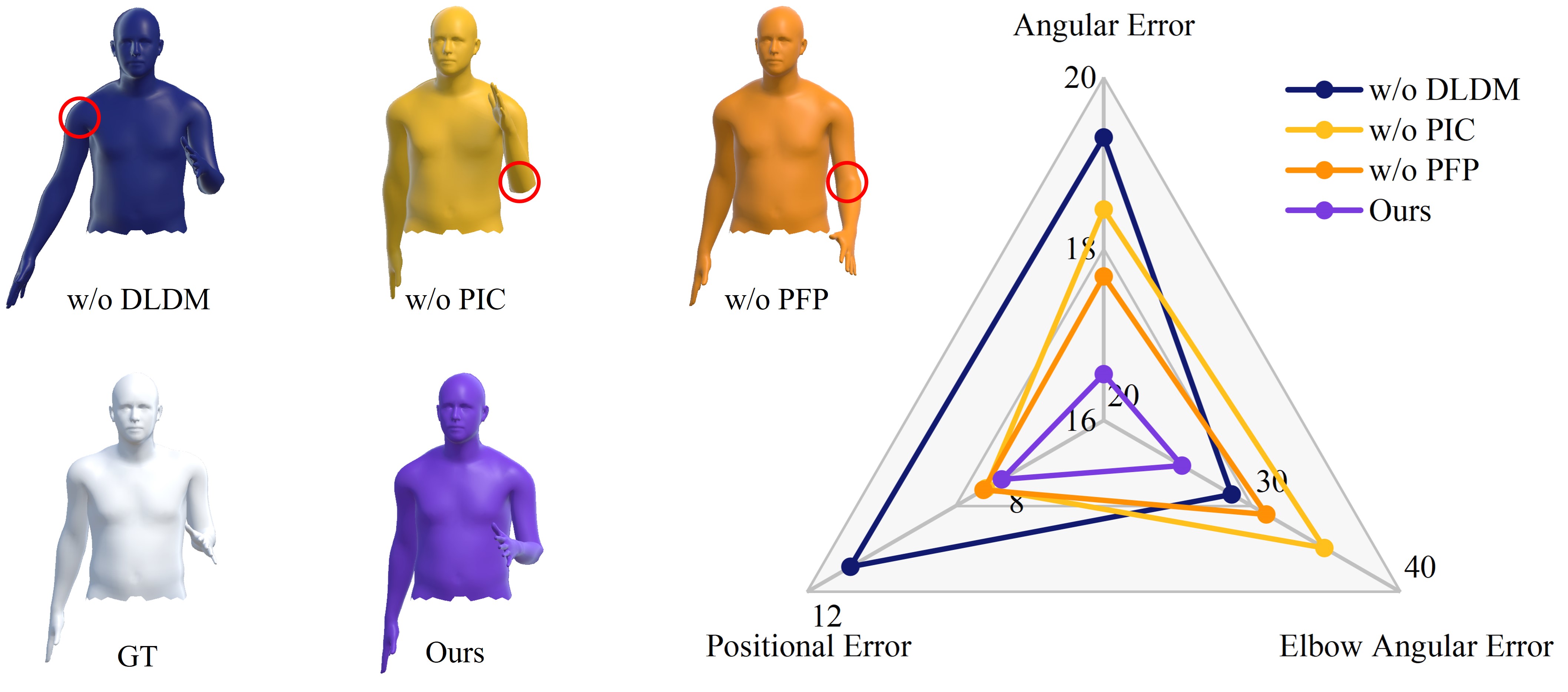}
\caption{The results of the ablation study show that the performance declines when any component of DLDM, PIC, or PFP is removed, indicating that all three components are essential.}
\label{fig:ablation}
\end{figure}

We evaluate the effectiveness of the proposed Displacement Latent Diffusion Model, Physical-informed Calibrator, and Pose Fusion Predictor in Fig. \ref{fig:ablation}. The results demonstrate that all three proposed techniques are necessary to achieve the best performance.

\begin{figure}[ht]
\centering
\includegraphics[width=1.0\linewidth]{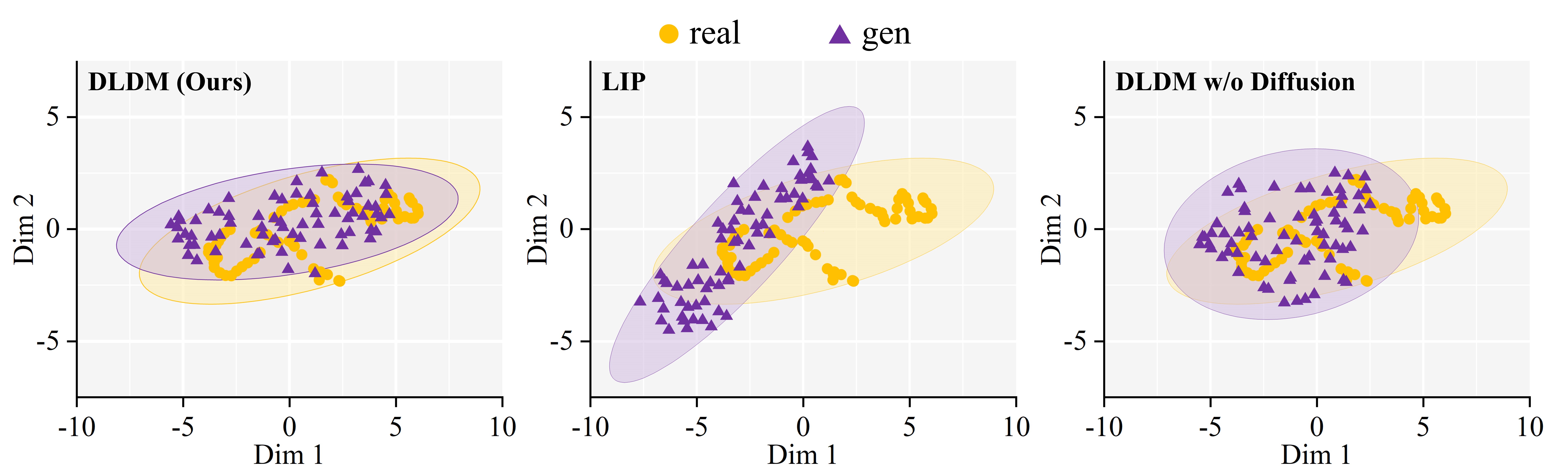}
\caption{Qualitative results of ablation study of DLDM. We use t-SNE to visualize IMU displacement generated by DLDM, LIP, and DLDM without Diffusion, respectively. The image shows: i) our DLDM generates more realistic IMU displacement data than LIP; ii) removing the proposed Diffusion Model will degrade the similarity with real-world data, highlighting the significance of Diffusion in DLDM.}
\label{fig:vs}
\end{figure}

Additionally, since our synthesis of IMU \textit{Real-time Displacement} is fundamentally a conditional generation task, we compared our DLDM generator with the generator of LIP and evaluated the effectiveness of Diffusion in DLDM. We assessed the richness and quality of the generated IMU displacement data quality using metrics such as FID, PSNR, and SSIM \cite{sara2019image}, as detailed in Tab. \ref{tab:fipvslip}. Besides, the generated IMU displacement data were compared using t-SNE dimensionality reduction, as illustrated in Fig. \ref{fig:vs}. These evaluations demonstrate the significance of Diffusion in DLDM and highlight the superiority of DLDM over LIP. The IMU displacement data generated by our DLDM offers better fidelity and diversity compared to LIP. This enhancement enables our approach to more effectively simulate real-time IMU displacements in practical applications.


\begin{table}[h]
\centering
\caption{Quantitative results of ablation study of DLDM. These results demonstrate the significance of Diffusion in DLDM and highlight the superiority of DLDM over LIP.}
\label{tab:fipvslip}
\begin{tabular}{lccc}
\toprule
Method & FID↓          & PSNR↑          & SSIM↑          \\ \midrule
LIP    & 1.54          & 11.26          & 0.73           \\
DLDM (w/o Diffusion)     & 1.48          & 10.26          & 0.70           \\
\textbf{DLDM (Ours)}       & \textbf{1.17} & \textbf{12.51} & \textbf{0.78}  \\ \bottomrule
\end{tabular}
\end{table}

\subsection{Real-time MoCap}
\label{sec:realtime}
\begin{figure}[ht]
\centering
\includegraphics[width=1.0\linewidth]{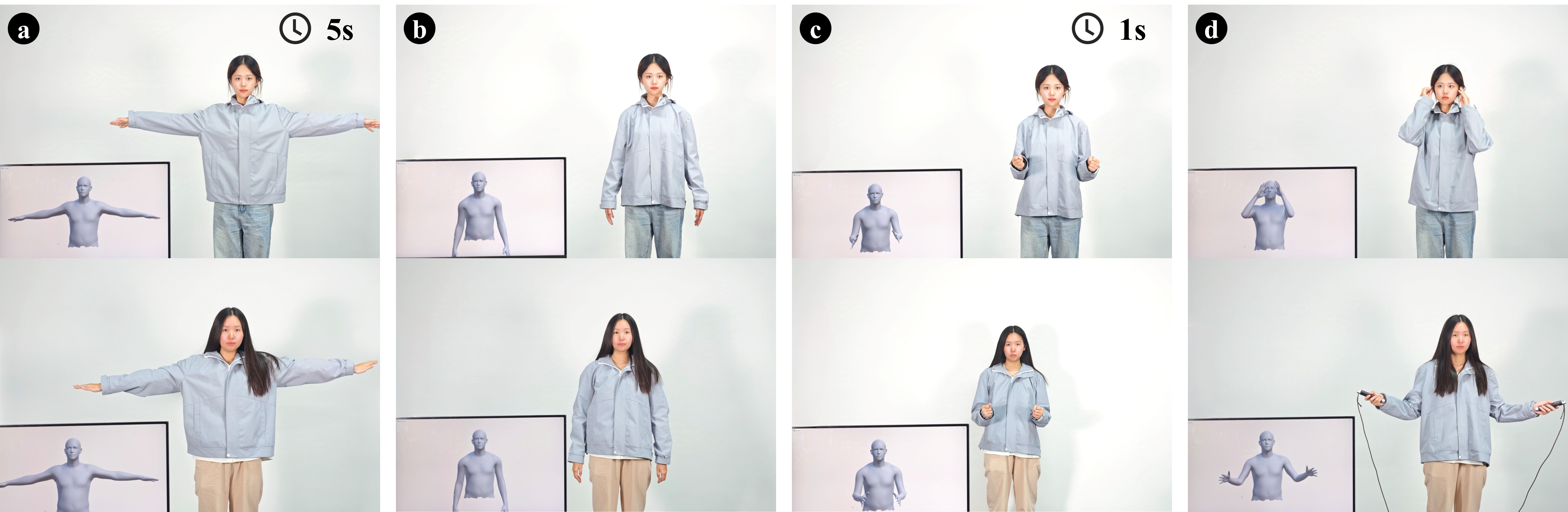}
\caption{Demonstration of the real-time MoCap visualization system. (a) a T-pose calibration (5 seconds); (b-c) an elbow flexion calibration (1 second); (d) real-time MoCap visualization result.}
\label{fig:live}
\end{figure}

We implemented a real-time MoCap visualization interface using Python and Unity (Fig. \ref{fig:live}). When the user dons our garment as common, the system initiates with a T-pose calibration (5 seconds) followed by an elbow flexion calibration (1 second). During operation, the system continuously receives real-time IMU and flex sensor readings. The IMU readings are calibrated and normalized using methods similar to those in TransPose\cite{yi2021transpose}, while the flex sensor readings are calibrated separately through the PIC. Both readings are then input into the PFP to achieve real-time motion capture outputs. Through the real-time MoCap visualization, we demonstrate the system’s convenience, comfort, and robustness in motion capture, particularly excelling in elbow movement tracking. A demonstration video of the real-time MoCap system is provided in the supplementary materials.

\section{Applications}
\begin{figure}[ht]
\centering
\includegraphics[width=1.0\linewidth]{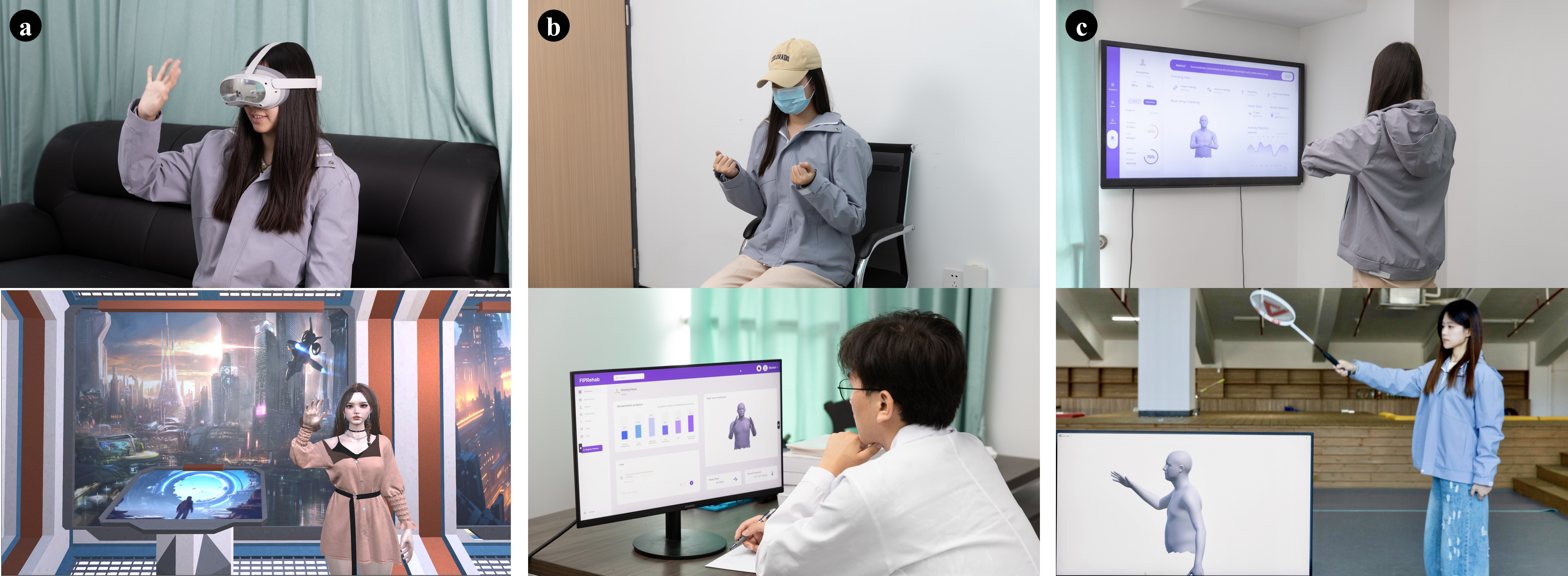}
\caption{Applications of our approach: (a) Metaverse, (b) rehabilitation, (c) fitness analysis.}
\label{fig:app}
\end{figure}

We demonstrate applications of our \textit{Clothes-based MoCap} system in various human-computer interaction scenarios, including virtual and augmented reality (VR/AR), rehabilitation, and fitness analysis (Fig. \ref{fig:app}), leveraging the system's robustness, accessibility, and comfort.

\subsection{Virtual \& Augmented Reality}
MoCap is essential for AR/VR, facilitating the creation of digital twins in the Metaverse and providing a realistic experience for users. Our garment-based system offers superior privacy protection compared to visual capture solutions, mitigating risks of information leakage associated with cameras. It also serves as a self-contained system, eliminating the need for external cameras and remaining unaffected by visual occlusion.  Additionally, unlike other wearable devices that often require tight-fitting, our loose-fitting system ensures ease of use and comfort while maintaining accurate and reliable motion capture. This allows users to effectively control their avatars in the virtual world (Fig. \ref{fig:app}a), unlocking new possibilities for immersive interaction in AR/VR applications.

\subsection{Rehabilitation}
Home rehabilitation has a significant need for ubiquitous and longitudinal monitoring. Our garment-based system is particularly well-suited for home rehabilitation monitoring because of its robustness, comfort, and seamless integration. Unlike conventional medical devices, the garment is a non-intrusive and lifestyle-friendly solution that patients can wear as regular garments continuously in everyday life. Through our system, healthcare providers can remotely and in real-time monitor patient progress and offer personalized treatment recommendations based on captured motion data (Fig. \ref{fig:app}b). This approach not only improves patient adherence during rehabilitation but also enhances comfort and privacy.

\subsection{Fitness Analysis}
In the domain of fitness and sports, MoCap is widely used to analyze techniques, prevent injuries, enhance performance, and tailor personalized training programs by providing detailed insights into movement patterns and biomechanics. Our \textit{Clothes-based MoCap} method provides an effective and practical approach for motion capturing in both indoor and outdoor activities, eliminating the need for complex setups or additional equipment. 
By collecting and analyzing motion data in real-time, the system can effectively assess user posture and activity levels, offering tailored exercise recommendations (Fig. \ref{fig:app}c). The combination of flex sensors and IMUs ensures reliable motion capture, aiding users in adjusting their training regimens for optimal results.

\section{Discussion}
\subsection{Clothes-based MoCap}
\paragraph{Advantages} Clothing provides an efficient medium for sensor fusion, which can significantly enhance motion capture accuracy without compromising user comfort. As a compelling exploration, FIP leverages both IMUs and flexible bend sensors, capitalizing on the strengths of each to improve motion capture precision, particularly in accurately measuring elbow joint angles. This approach could inspire researchers to incorporate a wide range of sensors, such as ToF, laser, radar, and ultrasound. More importantly, we emphasize that this fusion has the potential to revolutionize the limitations of previous sensor measurement paradigms, enabling biomechanical functions beyond simple joint tracking. For instance, some studies have utilized flexible strain sensors to measure the wearer's body dimensions, which, when integrated with IMUs, could allow for the simultaneous capture of both body shape and motion. Consequently, Clothes-based motion capture through sensor fusion holds significant promise for applications in fitness analysis, rehabilitation, and more.
\paragraph{Disadvantages} 
While integrating multiple sensors into clothing can significantly enhance motion capture accuracy, it also increases manufacturing complexity and costs. Additionally, the need to develop algorithms to address the unique displacement issues of each sensor presents further challenges in algorithm design.
\subsection{Limitations \& Future works}
\textit{On the hardware side}, although we have conducted comparisons with different users, the impact of clothing sizes and fabric materials on accuracy remains to be explored. \textit{On the software side}, although we have addressed the \textit{Primary Displacement} of flex sensors when users first don the clothing, as the \textit{Real-time Displacement} of the flex sensors located at the elbow is minimal in our garment. However, real-time correction of significant sensor displacements that may occur later (\textit{e.g.}, due to forceful tugging during movement) remains an open challenge. In the future, we plan to explore solutions to this issue to further enhance accuracy.

\section{Conclusion}
In this paper, we introduced Flexible Inertial Poser (FIP), a novel approach for achieving real-time, accurate pose estimation through the fusion of flex and inertial sensors. Our key innovation lies in the hardware prototype and software design of the sensor fusion garment. Specifically, FIP introduces a Displacement Latent Diffusion Model to synthesize sensor disturbance in various conditions from limited simulation data, as well as a Physiscs-informed Calibrator to rectify the\textit{ Primary Displacement} of flex sensors. Extensive experiments demonstrate that our method achieves robust performance across varying body shapes and motions, significantly outperforming SOTA approaches with a 19.5\% improvement in angular error, a 26.4\% improvement in elbow angular error, and a 30.1\% improvement in position error. As the first endeavor to integrate flex and inertial sensors in a MoCap garment, our work aims to inspire advancements in \textit{Clothes-based MoCap} within the HCI community.


\bibliographystyle{ACM-Reference-Format}
\bibliography{chi25}

\end{document}